\documentclass[fleqn,twoside]{article}
\usepackage{espcrc2}

\usepackage{graphicx}
\usepackage[figuresright]{rotating}

\newcommand{\AmS}{{\protect\the\textfont2
  A\kern-.1667em\lower.5ex\hbox{M}\kern-.125emS}}
\newcommand{\Zo}{{\mathrm Z}^0}
\newcommand{\eett}{\ensuremath{\mathrm{e^+ e^- \rightarrow \tau^+ \tau^-}}}
\newcommand{\eettg}{\ensuremath{\mathrm{e^+ e^- \rightarrow \tau^+ \tau^- \gamma}}}
\newcommand{\ggtt}{\ensuremath{\mathrm{\gamma \gamma \rightarrow \tau^+ \tau^-}}}
\newcommand{\Ztt}{\ensuremath{\mathrm{Z \rightarrow \tau^+ \tau^-}}}
\newcommand{\ttee}{\ensuremath{\mathrm{e^+ e^- \rightarrow e^+ e^- \tau^+ \tau^-}}}
\newcommand{\mmee}{\ensuremath{\mathrm{e^+ e^- \rightarrow e^+ e^- \mu^+ \mu^-}}}

\newcommand{\EE}{\ensuremath{\mathrm{e^+ e^-}}}

\def\aw{a^w_\tau}
\def\dw{d^w_\tau}
\def\at{a_\tau}
\def\dt{d_\tau}
\def\taurho{\tau^+ \rightarrow \rho^+ \nu_\tau}
\def\taue{\tau^- \rightarrow \rm{e}^- \nu_e \nu_\tau}

\def\etal{{\it et al.}}

\def\Journal#1#2#3#4{{#1} {\bf #2}, #3 (#4)}

\def\NPB{{\em Nucl. Phys.} B}
\def\NPO{{\em Nucl. Phys.}}
\def\PLB{{\em Phys. Lett.}  B}

\def\PRL{\em Phys. Rev. Lett.}
\def\PRD{{\em Phys. Rev.} D}

\def\RMP{{\em Rev. Mod. Phys.}}
\def\ZPC{{\em Z. Phys.} C}
\def\EPJ{{\em Eur. Phys. Jour.} C}

\title{Electromagnetic and Weak Moments of the $\tau$-Lepton \thanks{Talk given 
at the International Workshop on Tau Lepton Physics, Nara, Japan, Sept. 2004.}}

\author{W. Lohmann\address{DESY
         \\ 
        Platanenallee 6, 15738 Zeuthen, Germany}%
             }
       
\begin{document}

\begin{abstract}
The electromagnetic and weak dipole moments of the  $\tau$-lepton
have been measured by experiments at \EE colliders. Data samples of 
\eett, \eettg~and~\ttee~events collected at centre-of-mass energies between
10 and 200 GeV are used. No deviation from the Standard Model is found.
Limits on the moments are summarised from the most recent results.
\end{abstract}

\maketitle

\section{Introduction}
In the Standard Model (SM)~\cite{sm} leptons are pointlike fermions.
The magnetic moments are predicted with high precision 
by the theory and the electric dipole moments must be zero.
The 
moments of the electron are measured with high 
precision~\cite{m_ele}
and found in perfect agreement with the predictions.
For the muon magnetic moment a discrepancy of
2.7 standard deviations is found between recent measurements and the 
prediction~\cite{hertz},
giving rise to think about a possible
indication of physics beyond the SM. 
A deviation of the magnetic moments from the predictions of the theory 
would be a signal for a substructure or new interactions.
Furthermore, a measurement of non-zero values of the electric (electromagnetic or weak)
dipole moments reveals CP violation.
For the $\tau$-lepton, about 17 times heavier than the muon, only rough 
limits on the moments
were derived before the LEP era~\cite{petra}.
At LEP experiments,
measurements of the electromagnetic moments are done using the 
processes \eettg~and,
recently, \ttee. Also BELLE published a measurement 
of the electric dipole moment of the $\tau$-lepton
in \eett. 
Weak moments are measured from the process \eett at centre-of-mass energies near the 
$\Zo$ resonance by several experiments. The ALEPH experiment
published a new result using the full data statistics.
 A review of the
experimental results is given.

\section{Couplings of the $\tau$-lepton to the Photon and the $\Zo$}

\subsection{Electromagnetic Moments}

\begin{figure}[ht]
\includegraphics[width=7cm,clip]{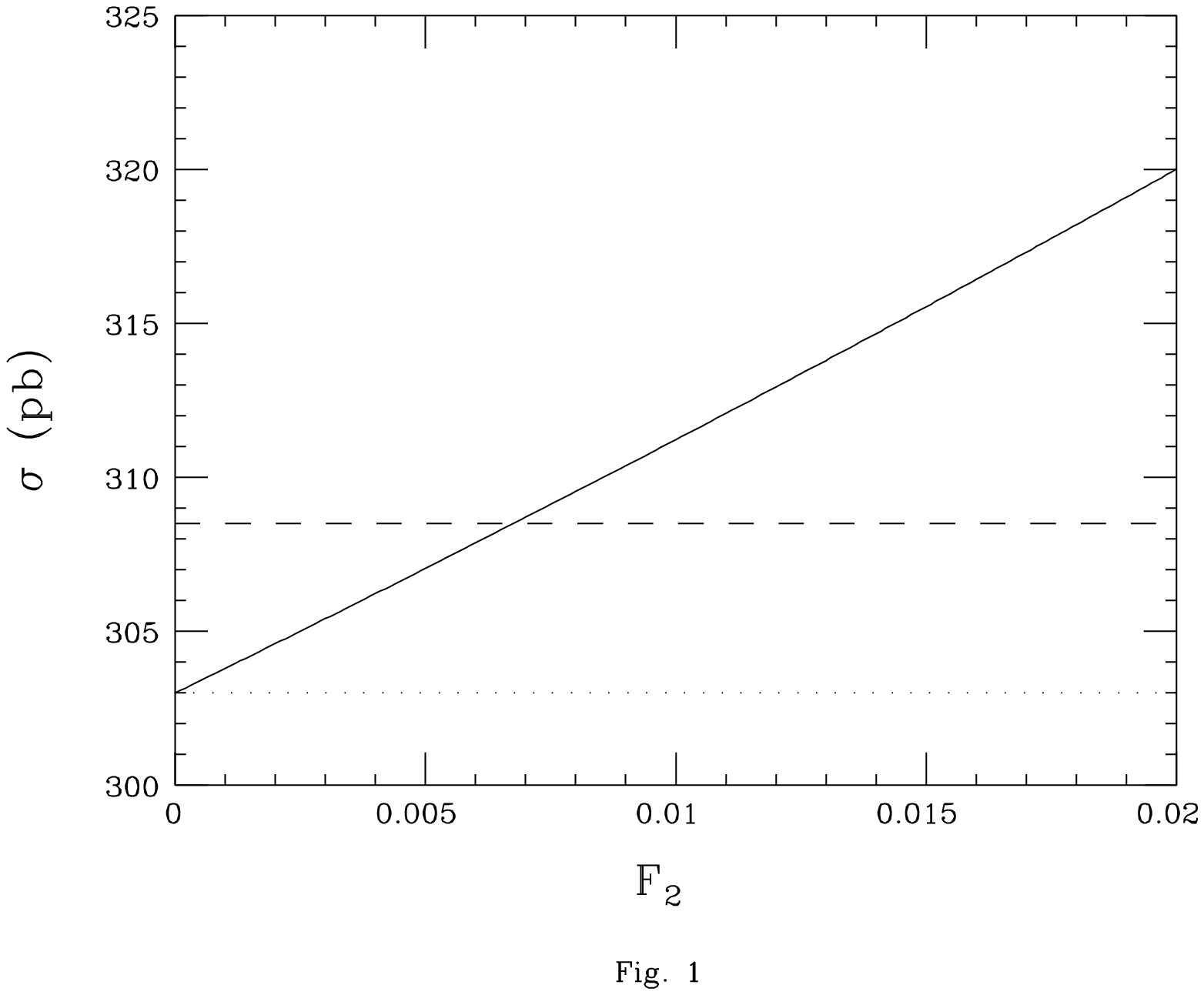}
\includegraphics[width=7cm,clip]{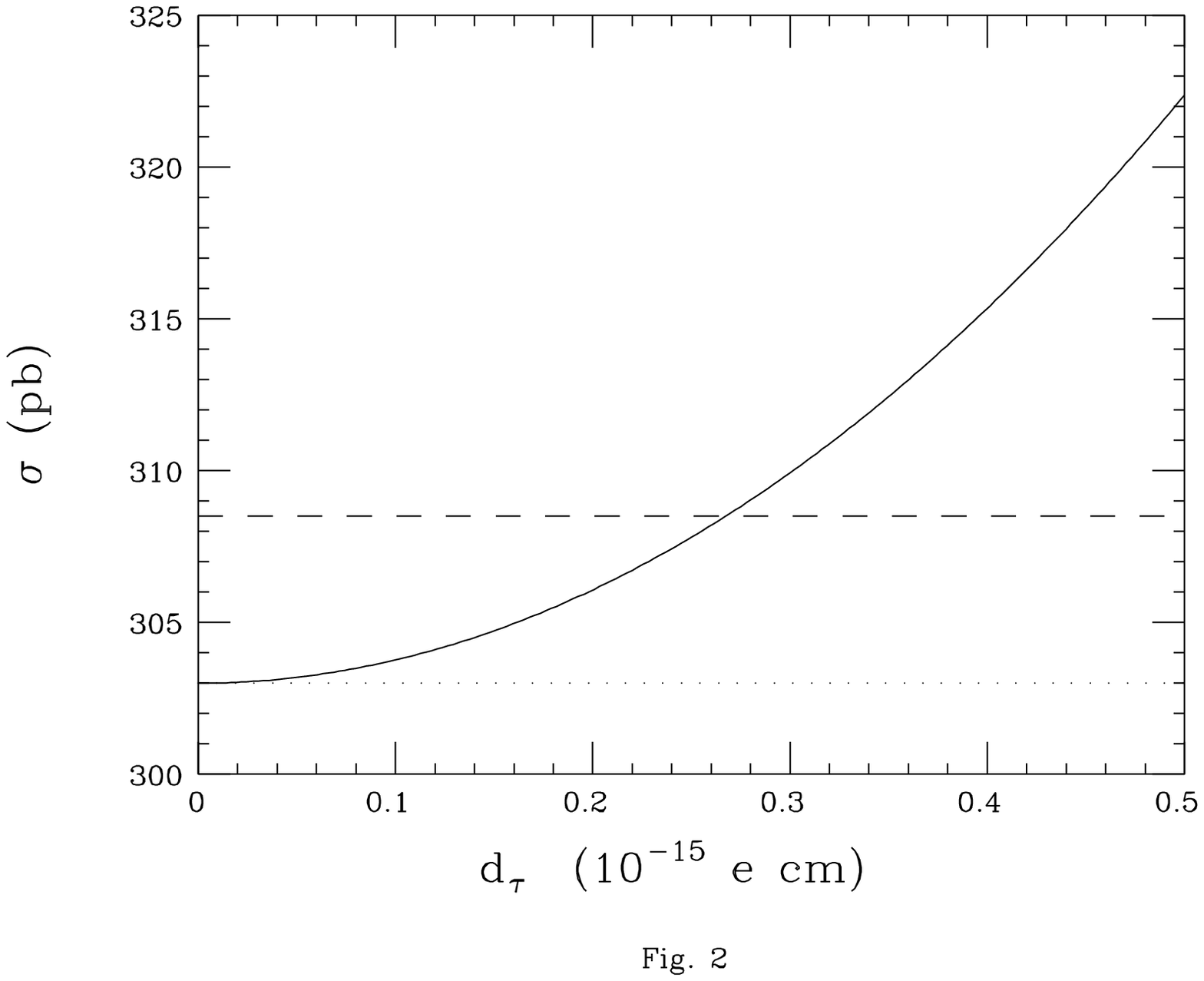}
\vspace{-.8cm}
\caption{The cross section of \ttee~as a function
of $F_2$ = $a_\tau$ (top) and as a function of  $d_\tau$ (bottom).}
\label{fig:ggcs}
\end{figure}
In general a $\tau$-lepton couples to a photon through its electric charge, 
magnetic moment and electric dipole moment. The Lorentz-invariant ansatz
is given by the following form~\cite{wein95},

\begin{eqnarray}
\Gamma^\mu &=& F_1(q^2)\gamma^\mu +  F_2(q^2)\frac{i}{2m_\tau}\sigma^{\mu\nu}q_\nu\\
           & & -  F_3(q^2)\sigma^{\mu\nu}\gamma^5q_\nu \nonumber.
\end{eqnarray}

The $q^2$ dependent form-factors, $F_i(q^2)$, have familiar interpretations for
 $q^2 = 0$: $Q_\tau$ = $e F_1(0)$, where $Q_\tau$ is the charge of the 
            $\tau$-lepton and  $e$ the unit charge, 
            $F_2(0)$ = $a_\tau$ is the anomalous magnetic moment, $a_\tau$ = 
            $(g_\tau-2)/2$, and
            $F_3(0)$ = $d_\tau/e$, where $d_\tau$ is the electric
dipole moment. In the SM $a_\tau$ is non-zero due to higher order diagrams
and is predicted to be $a_\tau^{SM}$ = 0.001 177 3(3)~\cite{samuel}.
A non-zero value of $d_\tau$ is forbidden by both P invariance and 
T invariance~\cite{barr}. Assuming CPT invariance, the observation of a non-zero value
of $d_\tau$ would imply CP violation.

$F_2(q^2)$ and $F_3(q^2)$ can be probed in
processes with $\gamma \tau \tau$
vertices. 
Hereafter I follow the convention $a_\tau$= $F_2$ and $d_\tau$ = $F_3 \cdot e$.

In the process \eett~at low energies the photon is virtual 
with $q^2$ = s, where s is the centre-of-mass energy.
Non-zero values of $a_\tau$ and $d_\tau$ contribute to the
cross section~\cite{petra}. 
Exploiting also the $\tau$ lepton decays
in \eett, additional information on $d_\tau$ can be obtained from triple
momentum and spin correlation observables~\cite{overmann}. The joint spin-density
matrix of the  $\tau$ leptons can be written
as
\begin{eqnarray}
\chi & =& \chi_{SM} + {Re(~d_\tau)} \cdot  \chi_{CP}^{Re} + {Im~(d_\tau)} \cdot \chi_{CP}^{Im} \nonumber \\
     &  &  + |d_\tau|^2  \chi_{CP}^{d^2}.
\end{eqnarray}
The term $\chi_{SM}$ results from the SM amplitudes, $\chi_{CP}^{Re}$ and $\chi_{CP}^{Im}$
are the interference terms between SM and CP violating amplitudes
and $\chi_{CP}^{d^2}$ is the CP even contribution
bilinear in $d_\tau$.
The term  $\chi_{CP}^{Re}$ is CP odd and T odd while $\chi_{CP}^{Im}$ is  CP odd and T even.

The coupling to real photons appears in \eettg~(final state bremsstrahlung).
Non-zero dipole moments enhance the production of high energy photons in
\eettg~\cite{riemann}. For small $a_\tau$ the effect is due to the
 interference between the SM and the anomalous amplitudes leading to a linear 
dependence in $a_\tau$. The electric dipole moment contributions
depend on the square of $d_\tau$,
allowing only the determination of its absolute
value. In this process $q^2$ is zero but 
the radiating $\tau$ is not on the mass shell.

Almost real photons are expected in the two-photon process, \ttee.
The cross section of \ttee~depends on a quartic polynomial in 
$a_\tau$~\cite{illana}. For 
small values of $a_\tau$ again only the linear term in $a_\tau$ contributes.
In case of $d_\tau$ the lowest order dependence is quadratic.
For illustration, Figure~\ref{fig:ggcs} shows the change of the total cross section
of \ttee~as a function of $a_\tau$ and $d_\tau$.

\subsection{Weak Dipole Moments}   

Weak dipole moments are introduced to the $Z \tau\tau$ vertex
in a similar way as for the electromagnetic ones in eqn.(1)~\cite{bernabeu}.
The quantity $F_2(q^2=m_Z^2)$ = $a_\tau^w$ is the anomalous weak magnetic 
dipole moment and  $F_3(q^2=m_Z^2)$ = $\dw$/$e$ is the weak 
electric dipole moment.
Both quantities are zero in the SM at Born level. Loop diagrams
result in $\aw$ = $-(2.10+0.61i) \cdot 10^{-6}$~\cite{ber1} and 
$\dw < 8 \cdot 10^{-34} e$cm~\cite{booth}, where $e$ is the unit charge.
These numbers are well beyond the sensitivity of current experiments.
However, some models beyond SM predict values up to 
$10^{-3}$ for $\aw$~\cite{gonzalez} and $10^{-19} e$cm for 
$\dw$~\cite{poulose}.  

The measurement of $\aw$ and $\dw$ in \Ztt
is possible from the transverse and normal polarisations 
of the
$\tau$-leptons~\cite{bernabeu}.
The full sensitivity is obtained if the $\tau$-lepton
direction of flight is reconstructed. This is possible
when both $\tau$-leptons decay semileptonically~\cite{kuhn}.    
In some measurements also the 
ansatz given in eqn.(2) is used.  

\section{Measurements of $\at$ and $\dt$}

\begin{figure}[h]
\includegraphics[width=7cm]{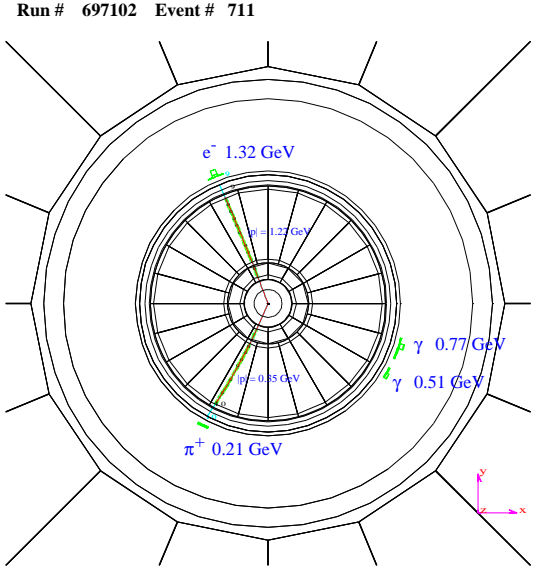}
\caption{An event display of \ttee~in the L3 detector. 
Two low momentum tracks point to local energy depositions
in the BGO calorimeter. In addition, two photons originating from a 
$\pi^0$ are seen. The event is identified as a 
$\taurho~$, $\taue$ final state.
}
\label{fig:l3event}
\end{figure}
 \begin{figure}[h]
\includegraphics[width=7cm,height=5cm]{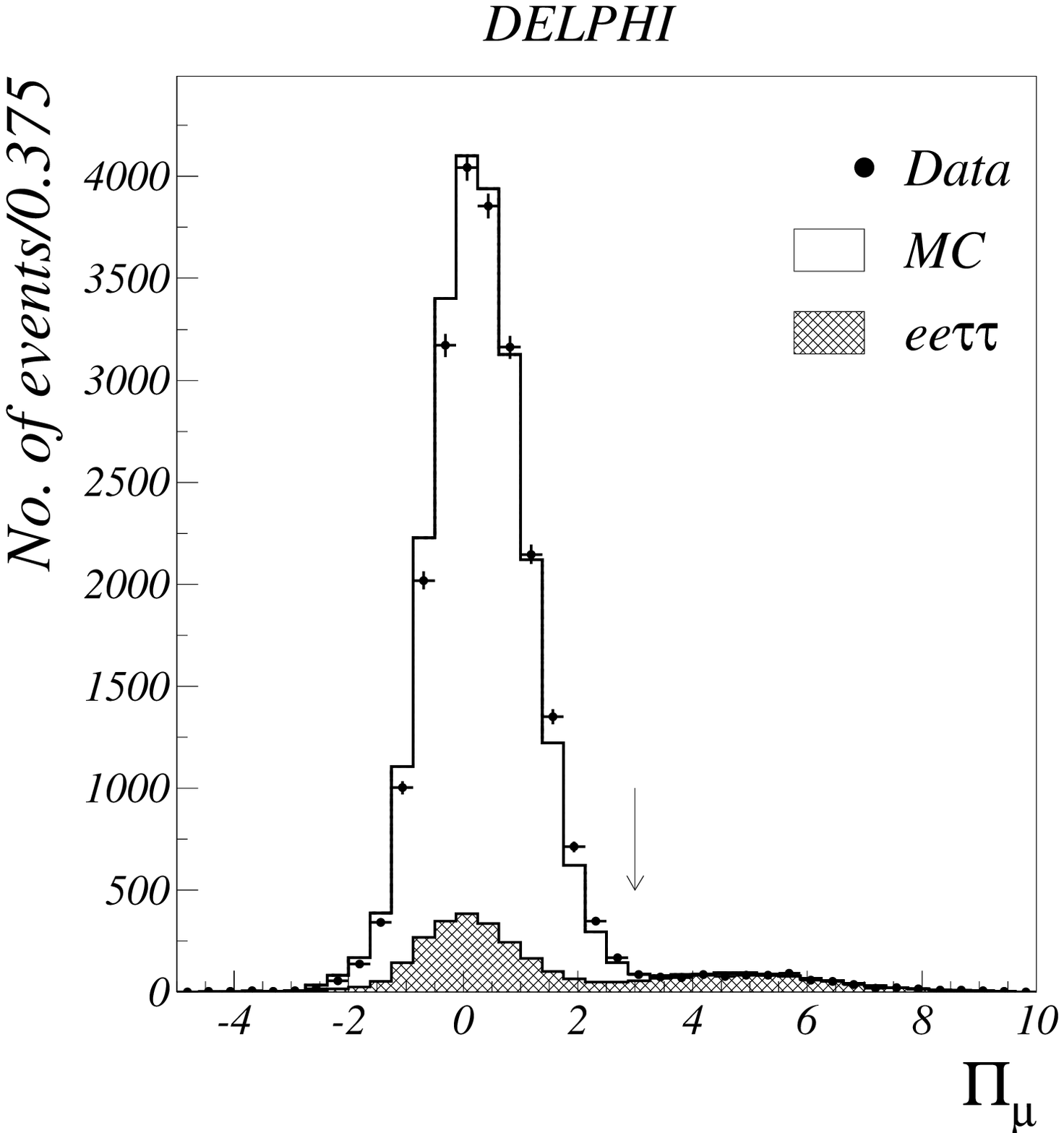}
\includegraphics[width=7cm,height=5cm]{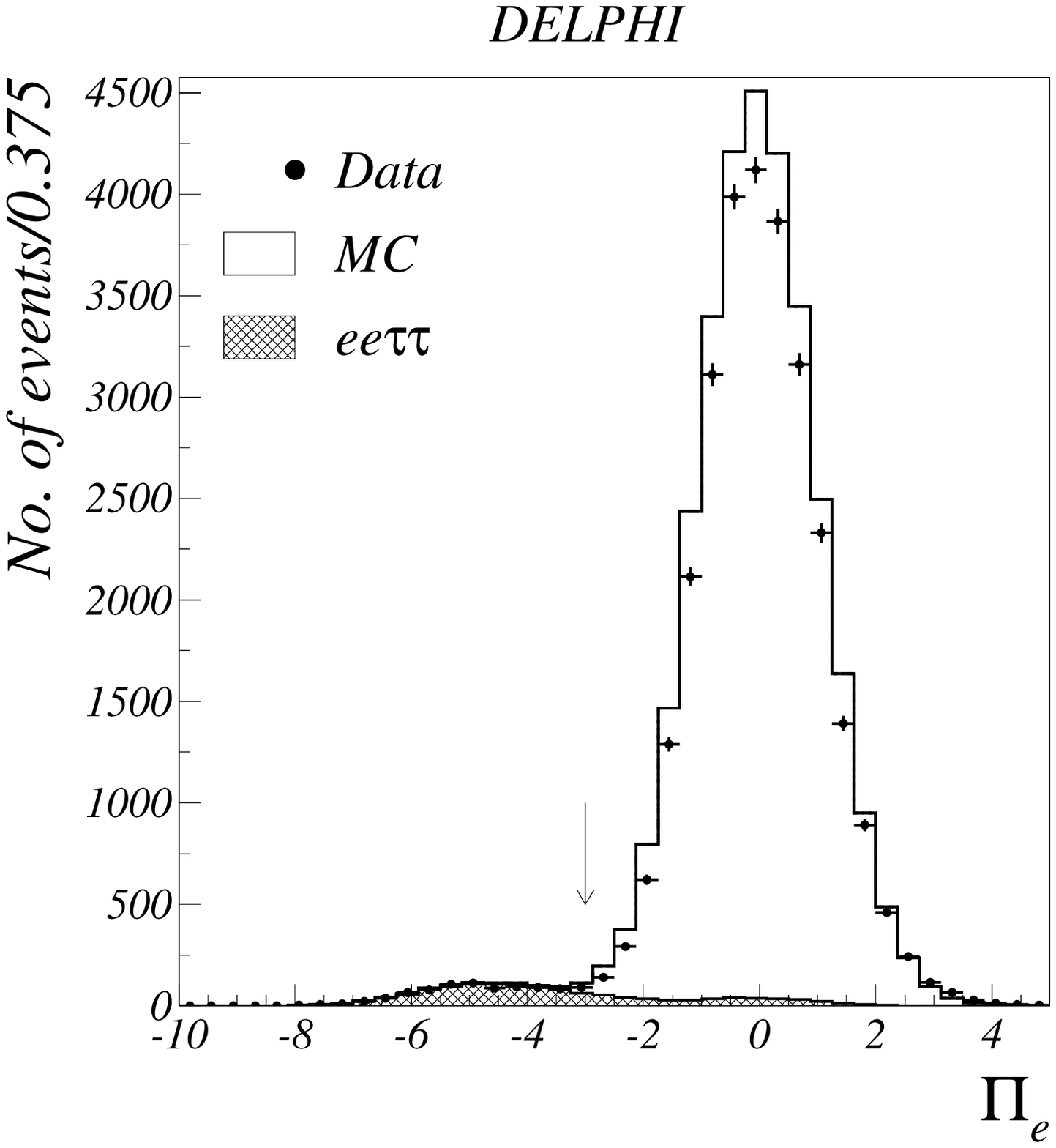}
\caption{The pull distributions for the muon hypothesis
(top) and the electron hypothesis (bottom)
for preselected two-photon events in the DELPHI experiment.
Using a proper combination of these informations,
one-prong 
$\tau^+\tau^-$ decays into an electron and a non-electron
are selected.
}
\label{fig:Dpull}
\end{figure} 
The first estimate of $\at$ was done using data from PETRA~\cite{petra}
at $q^2 \approx$  1000 GeV$^2$, resulting in
$\at \le 0.02$. Indirect limits were inferred
from the partial width $\Gamma(\Ztt)$~\cite{masso} measured at LEP,
yielding -0.002$\le \at \le$0.006 and $|d_\tau| \le 1.1 \cdot 10^{-17}e$cm.
The first attempt to derive a value at   
 $q^2$ = 0 was done in Ref.~\cite{grifols} using data from the L3 experiment at LEP.
The latter method was improved, on the basis
of the new calculations~\cite{riemann} and applied to the data
by the OPAL~\cite{opal1} and L3~\cite{L32} experiments. 
Using the
photon energy
spectrum and the distribution of the angle between the photon
and the nearest $\tau$-lepton,
%
a likelihood fit done by L3 results
to $\at=0.004\pm0.027\pm0.023$ and $\dt=(0.0\pm1.5\pm1.3)\cdot 10^{-16}
e$cm. 
Recently DELPHI~\cite{delphi} and L3~\cite{L3gg} 
analyzed the \ttee channel to determine these moments.
Both experiments used the data of LEP in the centre-of-mass
energy range between $\sqrt s$ = 180 GeV and 208 GeV,
corresponding to an integrated luminosity of 650 pb$^{-1}$
per experiment. Only events are used  
with non-detected electrons
and positrons, ensuring that $q^2 <$ 1 GeV$^2$.
The selection of \ttee events requires low deposited energy 
in the detectors and two non-coplanar low momentum tracks.
A typical event in the L3 detector is shown in
Figure~\ref {fig:l3event}.   

\begin{figure*}[htb]
\begin{center}
\includegraphics[width=6cm,height=5.5cm]{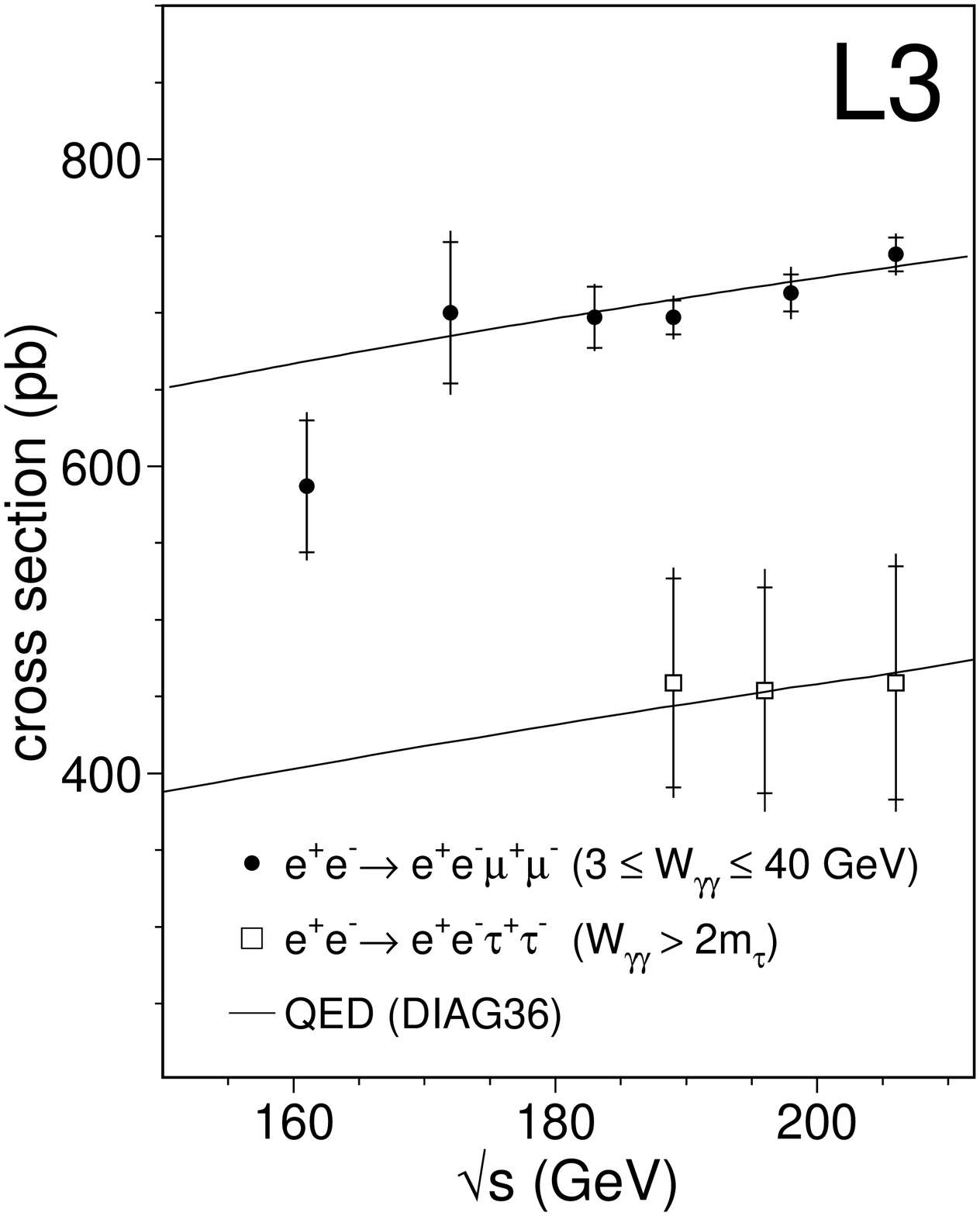}
\includegraphics[width=7cm,height=6.1cm]{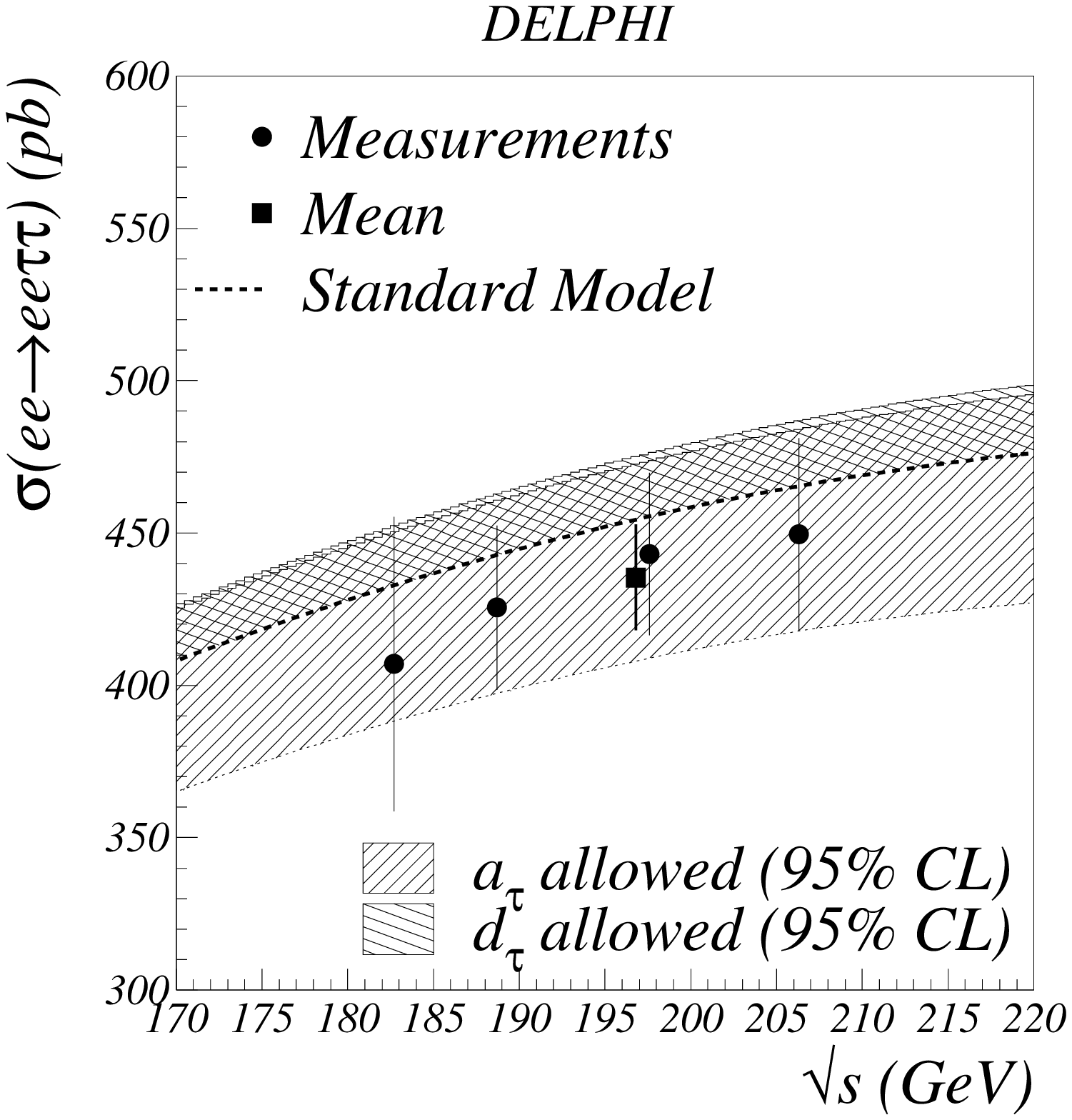}
\end{center}
\vspace{-.8cm}
\caption{The \ttee cross section as a function of the
centre-of-mass energy measured by the L3 (left) and
DELPHI experiments (right).
Also shown is the cross section for the process \mmee measured by L3. 
The solid lines in the L3 results are the expectation from the SM.
In the DELPHI result the dashed line corresponds to the SM prediction.
Also given by DELPHI is the one standard deviation 
range corresponding to the
fitted values of $\at$ and $\dt$ }
\label{fig:ggresults}
\end{figure*} 
L3 accepted events with one $\tau$-lepton decaying into an electron
and the other 
$\tau$-lepton decaying into a $\rho$-meson.
A sample of 351 events is selected. The 
selection efficiency is about 0.1\% and the background, mainly from other 
$\tau$-lepton decays, amounts to 30\%.
DELPHI selected events with one $\tau$-lepton decaying into an electron
and the other decaying in any one-prong final state but not an electron.
The identification of electrons and other particles is done using the 
ionisation along the track in the Time Projection Chamber (TPC).
A pull-quantity is defined for each track
by the normalised difference of the expected and measured energy loss
depositions for several particle hypotheses.
As an example, the pull for the muon hypothesis
and the electron hypothesis is shown
in Figure~\ref {fig:Dpull}.
Combining the pull information a sample of 2390 \ttee events is selected 
from a large two-photon event sample. The efficiency is of about 0.8\% and
the background of about 12\% results mainly from other two-photon final states.

The measured cross sections, 
shown in  Figure~\ref {fig:ggresults}
as a function of the centre-of-mass energy,
are compared to the expectation from the SM and a fit is performed
with 
cross section predictions including $\at$ and $\dt$ as free parameters
\footnote{Fits are done for each parameter separately, keeping the other 
fixed at the SM value.}.

Since the measurements are in reasonable agreement with the SM predictions,
there is no indication for non-SM values of the $\tau$ lepton moments. 
For the channels \eettg~and \ggtt, respectively, 
OPAL and L3 published only the limits
\mbox{
-0.068$<\at<$0.065}, \mbox{-3.7$<\dt<$3.7 ($ 10^{-16}e$cm)}
and  \\
\mbox{-0.107$<\at<$0.107},
\mbox{-1.14$<\dt <$1.14 ($ 10^{-15}e$cm).} 
The results from DELPHI and L3 measurements for the channels
\ttee~and  \eettg~are summarized in
Table~\ref{table:tabtau}.
\begin{table}[htb] 
\caption{
The values of $\at$  and $\dt$ measured by DELPHI in \ttee~and by L3 
in \eettg.
}
\label{table:tabtau}
\begin{center}
  \begin{tabular}{|l|c|c|} \hline
   Experiment   &  $\at$ & $\dt$(10$^{-16}e$cm) \\ \hline
   DELPHI   & -0.018$\pm$0.017   & 0.0$\pm$2.0  \\ 
   L3      & 0.004$\pm$0.035   & 0.0$\pm$2.0  \\ 
    \hline
  \end{tabular} 
\end{center}
\vspace{-.7cm} 
\end{table}

Combining the latter two results
one obtains \mbox{
$\at$ = -0.013$\pm$0.015} and 
 \mbox{$\dt$ = (0.0$\pm$1.7)$\cdot 10^{-16}e$cm}.

The process \eett~at $\sqrt s \approx$10 GeV was used by the 
ARGUS~\cite{argus}  and BELLE~\cite{belle}  collaborations
to search for CP violation due to a non-zero value of 
$\dt$. 
The integrated luminosity is 291 pb$^{-1}$  and 29.5 fb$^{-1}$
for the ARGUS and BELLE experiments, respectively. Both experiments
analyzed almost all one-prong decay channels, and clearly the 
sensitivity of BELLE is superior.
The result of the BELLE analysis is: $Re(\dt)$ = (1.15$\pm$1.70)
$\cdot$10$^{-17} e$cm and 
$Im(\dt)$ = (-0.83$\pm$0.86)$\cdot$10$^{-17} e$cm.

\section{Measurements of $\aw$ and $\dw$}

\begin{figure}[htb]
\includegraphics[width=8.5cm]{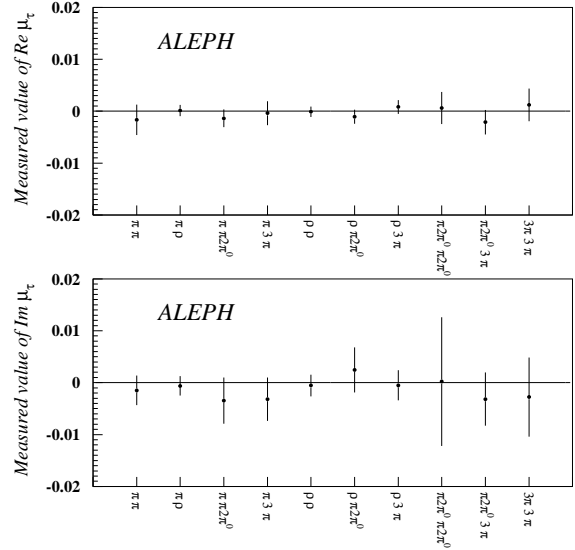}
\vspace{-1.2cm}
\caption{Results on $\aw$ (denoted here as $\mu_\tau$) for the
decay modes analyzed by the ALEPH experiment. Statistical and systematic
uncertainties are included.
}
\label{fig:aleph}
\end{figure}

Studies of $\dw$ were done first by OPAL~\cite{opal2} searching for CP 
violation using CP odd momentum tensors~\cite{nachtmann}. A review
on this and other results from LEP experiments was given at a previous
$\tau$ physics workshop~\cite{wermes}. The first measurement of $\aw$
was published by L3~\cite{L33} by measuring normal
and transverse polarisations~\cite{bernabeu} in \eett.
Exploiting these quantities in more detail,
ALEPH~\cite{aleph} published recently its first measurement of  $\aw$ and $\dw$
from a common fit to the full statistics recorded at LEP running on the Z,
corresponding to an integrated luminosity of 155 pb$^{-1}$.
One- and three-prong final states are used. 
The selection efficiency is about 50\% with 
background fractions, depending on the decay channel,
between 14\% and 42\%.
As an example, Figure~\ref{fig:aleph} shows the results of ALEPH for the real
and imaginary part of $\at$ for the decay modes analyzed.

The results combining all 
decay modes are
\mbox{$Re(\aw)$ = (-0.33$\pm$0.49)$\cdot 10^{-3}$,}
\mbox{$Im(\aw)$ = (-0.99$\pm$1.01)$\cdot 10^{-3}$,} \\
\mbox{$Re(\dw)$ = (-0.59$\pm$2.49)$\cdot 10^{-18} e$cm} and
\mbox{$Im(\dw)$ = (-0.45$\pm$5.67)$\cdot 10^{-18} e$cm}.
The quantities $Re(\aw)$ and $Im(\aw)$ are measured with much 
better precision than before.
A summary of all measurements of $\dw$ is given in Table~\ref{table:dtab}. 
\begin{table}[htb]
\caption{
The measured values of $Re(\dw)$ and $Im(\dw)$ as obtained
by the ALEPH, DELPHI and OPAL experiments.
}
\label{table:dtab}
\begin{center}
  \begin{tabular}{|l|c|c|} \hline
   Experiment   &  $Re(\dw)$ & $Im(\dw)$\\ 
                &  (10$^{-18}e$cm)&  (10$^{-17}e$cm)   \\                   \hline
   ALEPH    & -0.59$\pm$2.49   & -0.05$\pm$0.57  \\ 
   DELPHI   & -2.5$\pm$2.8     & -0.63$\pm$1.00   \\
   OPAL     &  0.72$\pm$2.47   &  0.35$\pm$0.58   \\
    \hline
  \end{tabular} 
\end{center}
\vspace{-.7cm}
\end{table} 
Averaging these values leads to $Re(\dw)$= (-0.65$\pm 1.49)\cdot 10^{-18} e$cm
and $Im(\dw)$ = (0.04$\pm$0.38)$\cdot 10^{-17} e$cm, or to 95\% CL limits of
($-3.56 < Re(\dw) < 2.26) (10^{-18}e$cm) and ($-0.69 < Im(\dw) < 0.77) (10^{-17} e$cm).   
\section{Summary}
The most recent measurements of the $\tau$-lepton 
electromagnetic and weak moments are in agreement with the Standard Model
predictions. No hint for new physics is observed. 
The accuracy is not sufficient to test the SM loop corrections.
In case that experiments published measured values of similar
accuracy, given in Table~\ref{table:tabtau} and Table~\ref{table:dtab},
the results are combined.
Otherwise the most
precise measurements are taken.
Limits
are obtained at 95\% C.L. for the electromagnetic moments,   \\
\begin{eqnarray}
-0.042 < &\at& < 0.016  \nonumber \\
-2.2   < &Re(\dt)& < 4.5~(10^{-17}e{\rm{cm}})\nonumber   \\ 
-2.5   < &Im(\dt)& < 0.8~ (10^{-17}e{\rm{cm}})\nonumber    
\end{eqnarray}
and for the weak moments:
\begin{eqnarray}
0.00114 < &Re(\aw)& < 0.00114   \nonumber \\
0.00265 < &Im(\aw)& < 0.00265   \nonumber \\
-3.56   < &Re(\dw)& < 2.26~(10^{-18}e{\rm{cm}})\nonumber   \\ 
-0.69   < &Im(\dw)& < 0.77~(10^{-17}e{\rm{cm}}).\nonumber    
\end{eqnarray}

\section{Acknowledgments}
I thank 
Francisco Matorras and Daniel Haas for the support given to me
to prepare this summary and Fred Jegerlehner
and Gunnar Klaemke for critical reading of the 
article. Klaus M\"onig I thank for helpful suggestions. 
Also I would like to thank H. Hayashii for the excellent organisation
of this workshop providing a creative and pleasant atmosphere.

\end{document}